\documentstyle[preprint,prl,aps]{revtex}
\begin{document}
\draft 
\title{Results from a High-Sensitivity Search for Cosmic Axions}
\author{C.~Hagmann, D.~Kinion, W.~Stoeffl, and K.~van Bibber}
\address{Lawrence Livermore National Laboratory \\
         7000 East Ave., Livermore, CA~~94550}
\author{E.~Daw, H.~Peng, and L.J~Rosenberg}
\address{Department of Physics and Laboratory for Nuclear Science\\
         Massachusetts Institute of Technology\\
         77 Massachusetts Ave., Cambridge, MA~~02139}
\author{J.~LaVeigne, P.~Sikivie, N.S.~Sullivan, and D.B.~Tanner}
\address{Department of Physics, University of Florida\\
          Gainesville, FL~~32611}
\author{F.~Nezrick}
\address{Fermi National Accelerator Laboratory\\
       Batavia, IL~~60510-0500}
\author{Michael S. Turner}
\address{Theoretical Astrophysics, Fermi National Accelerator Laboratory\\
Batavia, IL~~60510-0500}
\address{Departments of Astronomy \& Astrophysics and Physics,
Enrico Fermi Institute\\
The University of Chicago, Chicago, IL~~60637-1433}
\author{D.M.~Moltz and J.~Powell}
\address{Lawrence Berkeley National Laboratory\\
         1 Cyclotron Rd., Berkeley, CA~~94720}
\author{N.A.~Golubev}
\address{Institute for Nuclear Research of the Russian Academy of Sciences\\
          60th October Anniversary Prospekt 7a\\
          117 312 Moscow, Russia}
\author{}\author{}\author{}
\author{Accepted for Publication in Physical Review Letters}
\maketitle
\pagebreak[4]
\begin{abstract}
We report the first results of a high-sensitivity ($\sim 10^{-23}$~W)
search for light halo axions through their conversion to microwave
photons.
At 90\% confidence we exclude a KSVZ axion of mass
$2.9\times 10^{-6}\,$eV
to $3.3\times 10^{-6}\,$eV
as the dark matter in the halo of our Galaxy.
\end{abstract}


The dynamics of galaxies and of clusters of galaxies, as well as
their peculiar motions, imply that most of the mass of the Universe
is in an unseen form, called `dark matter'.
The amount of dark matter inferred is at least 20\% of the critical density,
and likely much more \cite{DM:review}.
Because the synthesis of the light elements in the big bang
restricts baryons to contribute no more than 10\% of the critical density,
a large nonbaryonic component is required.  The development of
structure in the Universe -- galaxies, clusters, and superclusters --
and the anisotropies of the cosmic
background radiation also support this conclusion.

The axion is a well-motivated particle dark matter candidate,
arising in models where the strong-CP problem is solved by
the Peccei-Quinn mechanism\cite{Kim:Cheng}.
The axion mass is constrained by laboratory experiments
and astrophysical limits to lie between 10$^{-6}\,$eV and
10$^{-3}\,$eV, with lower masses preferred if axions provide the bulk of
the critical density \cite{PR:Turner}.

If the dark matter is `cold' (small velocity dispersion),
as is indicated by studies of structure formation,
galactic halos are comprised primarily of
cold dark matter particles. Because dark matter 
axions were produced in a coherent
process in the early universe, they are cold \cite{axionguys}. 
Modeling of the Milky Way galaxy
indicates a local halo density $\rho_{halo}$
of $0.45$~GeV\,cm$^{-3}$
(about $7.5 \times
10^{-25}$~g\,cm$^{-3}$)\cite{Astro:Turner},
that implies an enormous
local density
${\cal O}(10^{14}\hbox{\rm cm}{^{-3}})$ of axions
if they are the dark matter. 
The velocity distribution of dark
matter particles is expected to be
approximately Maxwellian, with a dispersion of
$\langle \beta^2 \rangle^{1/2} \simeq 270\,$km/sec
\cite{PRD:Turner}.
There could also be narrow peaks in the velocity
distribution from dark matter particles which have recently fallen into the
galaxy and have yet to thermalize\cite{high:Sikivie}.
Because of its two-photon coupling,
${\cal L}_{a\gamma\gamma}$ $=$ $-g_{a\gamma\gamma} a\vec E \cdot\vec B$,
an axion can convert
to a single photon in the presence of a magnetic field
\cite{Sikivie:Krauss}.  Here $g_{a\gamma\gamma}= g_\gamma\alpha/\pi f_a$,
$f_a$ is the axion decay constant, the axion mass $m_a \simeq 6 \mu{\rm eV}\,
({10^{12}{\rm GeV}/f_a})$, and $g_\gamma$ is
a model-dependent coefficient of order unity.
In two popular models of the axion,
$g_\gamma = -0.97$ (KSVZ) and $0.36$ (DFSZ) \cite{Kim:Cheng}.

In a static magnetic field, the energy of the photon equals
that of the converted axion:
$E_\gamma = E_a = m_a + m_a\beta^2/2 = m_a (1 + {\cal O}(10^{-6}))$.
The conversion process is resonantly enhanced in a
high-Q cavity with resonant frequency $f_0$ tuned to $E_\gamma$, with power
given by\cite{Sikivie:Krauss}
\begin{eqnarray}
P = \left ({\alpha\over\pi} {g_\gamma\over f_a}\right )^2 
V\, B_0^2 
\rho_a C {1\over m_a} \hbox{Min}(Q_L,Q_a)
\end{eqnarray}
where $V$ is the volume of the cavity, $Q_L$ is the
loaded quality factor of the
cavity, $B_0$ is the central magnetic  field strength,
$\rho_a$ is the local axion density, and  
$1/Q_a\sim10^{-6}$ is the width of the axion energy
distribution.
The mode-dependent form factor $C$ is of order unity
for the  ${\rm TM}_{010}$ mode used
in our search and falls off rapidly for higher order modes.
For the parameters of this experiment and the KSVZ model,
$P\sim5 \times10^{-22}$~W.

Because the axion mass is unknown, the cavity resonant frequency
must be tuned.
When the ${\rm TM}_{010}$ resonant frequency is close
to the axion mass, the conversion of axions to
photons produces a narrow peak of fractional width
$\sim 10^{-6}$ in the cavity power spectrum.
The noise background is characterized by an effective
system temperature $T_s = T_c + T_a$,
where $T_c$ is the cavity physical temperature
and $T_a$ is the amplifier noise temperature.

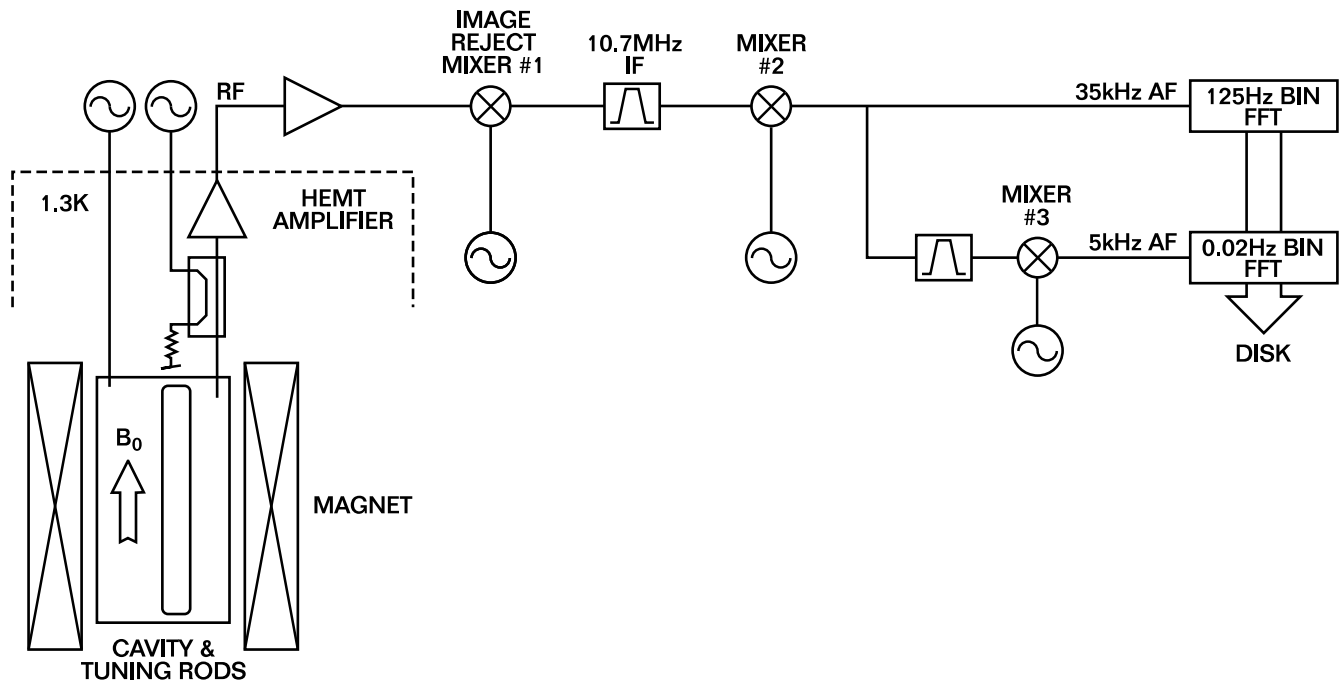
\begin{figure}
\caption{\leftline{Schematic diagram of the axion detector.}}
\label{sketch}
\end{figure}

Figure\ \ref{sketch}\ is a schematic diagram of the axion
detector, which is
located at LLNL\cite{NPB:Hagmann}.
The magnet is a superconducting solenoid of 7.6~T central field.
The cylindrical cavity (50~cm i.d., 100~cm long) is constructed
of stainless steel plated with copper
and subsequently annealed.
The temperature of the cavity is $T_c\sim$1.3K.
The resonant frequency $f_0$ of the empty
cavity is 460~MHz.
The unloaded Q, including losses
in the tuning rods, is $\sim$200,000, the limit set
by the anomalous skin depth of copper.
Two copper tuning rods, each 8~cm in diameter,
run the full length of the cavity.
The cavity is tuned by moving the rods
radially between the wall and center.
The cavity is normally evacuated, but can be
filled with liquid helium to shift the frequency from
the vicinity of mode crossings.
There are two coupling ports in the top of the cavity,
one weakly coupled and the other of variable coupling
strength. Power is extracted from the variable port through a
50~$\Omega$ transmission line, which is 50~cm long, to a directional
coupler and amplifier chain.
The weakly coupled port and the 30dB directional coupler
provide for transmission and reflection measurements
of cavity parameters.
The coupling at the variable port is adjusted to near critical.

The first- and second-stage amplifiers are balanced
GaAs HEMT devices built by the
National Radio Astronomy Observatory (NRAO).
They are cooled to the cavity temperature and are characterized by
noise
temperature $T_a$ $\sim$ 4.5K, power gain of
G $\sim$ 17dB, and power reflection
coefficient from the input of $\sim$-30dB\cite{daw}.
The amplifier noise temperature is measured by
varying the physical temperature of the cavity and
extrapolating the amplifier output to zero
physical temperature.

After 35dB of further amplification at room temperature,
the signal is down-converted to 10.7~MHz by an
image-rejection mixer.
An 8-pole crystal filter sets the 30~kHz
measurement bandwidth and prevents
image power from entering the second mixing stage. The
signal is then down-converted a second time, in effect shifting the cavity
resonant frequency to 35~kHz.

A commercial FFT spectrum analyzer then generates
the `medium-resolution' power spectrum.
During each 80-second run, 10,000 sub-spectra are measured and
averaged, resulting in a 400 point, 125~Hz/point power spectrum.
This is well-matched
to a search for the Maxwellian component of the halo,
which should be about 6 channels wide.

The analog output is also applied to a 6-pole filter
followed by a third mixing stage centering the cavity
resonant frequency at 5~kHz.
This signal is processed by a commercial ADC/DSP PC board,
yielding the `high-resolution' power spectrum.
There is no averaging,
but rather one 250,000 point,
0.02~Hz/point power spectrum is generated.
This is well matched to a search
for fine structure having fractional width $\sim {\cal O}(10^{-11})$
or less in the power spectrum.
If any appreciable fraction of the axions are in a narrow-velocity
line, it would be detected with high signal-to-noise ratio.
A cesium clock serves as the frequency reference
for both receivers.

After each 80-second run, the cavity frequency $f_0$ is tuned
upwards by 2~kHz, the loaded cavity quality factor Q$_L$ is measured,
and another run
initiated.  The dead-time associated with tuning and measuring cavity
parameters is about 4 seconds per run.
The form factor $C$ is calculated by a computer simulation of the
cavity at each frequency.
Each frequency range is swept out, then
the procedure is repeated at least twice more.
Regions in frequency where TE or TEM modes cross the TM$_{010}$ mode
are examined by filling the cavity with liquid helium, thus
shifting the mode-crossing frequency down by 3\%.
The overall live-time of the experiment has been well over
90\% since the beginning of data-taking.
Of the approximately $4.2 \times 10^5$ spectra
recorded for this analysis, 6058 were
eliminated due to anomalously large cavity frequency shifts
or pressure jumps, typically related to
cryogen fills or other disturbances.

During off-line data processing, the middle 200 frequency bins
of each 400 point spectrum are divided by the 8-pole
crystal filter response.
The resulting power spectrum varies slowly
with frequency by $\sim$1dB across the spectrum
due to noise from sources in the amplifier propagating
backwards along the transmission line and reflecting
from the cavity coupling.
To obtain a flat, corrected power spectrum, we
remove this slow variation using a 5-parameter
equivalent circuit model.

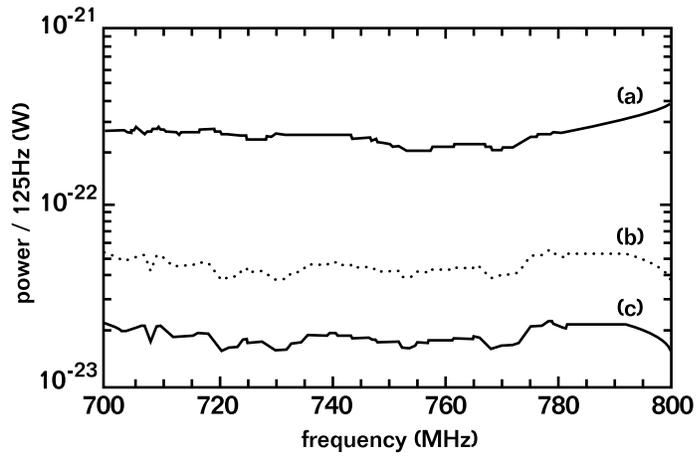
\begin{figure}
\caption{(a) The expected signal from KSVZ axions.
(b) The noise background for an axion signal distributed
over 6 channels.
(c) The noise background for an axion signal
confined to 1 channel.}
\label{snr}
\end{figure}

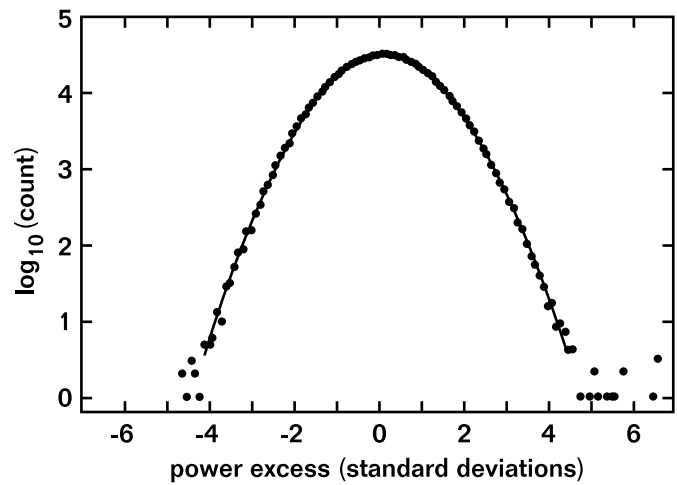
\begin{figure}
\caption{The distribution of the deviation of single channel noise power
from the thermal mean (in units of standard deviation $\sigma$),
in each 0.1~$\sigma$
interval, for all the data taken. The curve shows
the theoretical expectation (thermal Gaussian noise treated as input
to the analysis chain).
The slight deviation from Gaussian shape
is introduced by the filter response.
}
\label{gaussian}
\end{figure}

Because the typical 2~kHz tuning step is smaller than the
30~kHz crystal filter bandwidth, and because three or more
sweeps are made over the whole frequency range, each
125~Hz bin appears in at least 45 spectra.
These spectra are linearly combined with a weighting that
accounts for the $B_0$, $T_s$, $C$, $Q$, $f-f_0$
appropriate for each.
The result is a single spectrum of nearly $10^6$ points
between 701-800~MHz.
Figure\ \ref{snr}\ shows (a) the expected signal from
KSVZ axions,
(b) the noise background were the axion signal
distributed over 6 channels,
and (c) the noise background were
the axion signal
confined to 1 channel.
Figure\ \ref{gaussian}\ shows the deviation of the single channel power
from the thermal mean for all the data taken.
The distribution is consistent with thermal (Gaussian) noise out to 5$\sigma$.
This is an important validation of our understanding of the experiment.

Candidates for further examination are those single 125~Hz
channels with a 3.3$\sigma$ power excess (538 candidates), or
the sum of any 6 adjacent 125~Hz channels with a
2.25$\sqrt6\,\sigma$ power
excess (6535 candidates).
These candidates were then rescanned to the same power
sensitivity as the first spectrum.
Of the 6-channel (1-channel) candidates, 23 (6) persisted,
{\it i.e.,} appeared independently in the rescan.
Of those, 8 (4) were at the same frequency as known external RF sources.
The remainder developed power in the tails of the cavity Lorentzian 
instead of near the peak as would the axion.
This behavior is expected for external interference
introduced through the calibration ports and reflecting
off the cavity input back into the amplifier.
The few persistent candidates were rescanned after
terminating calibration lines leading into
the cavity; no candidates survived all scans.

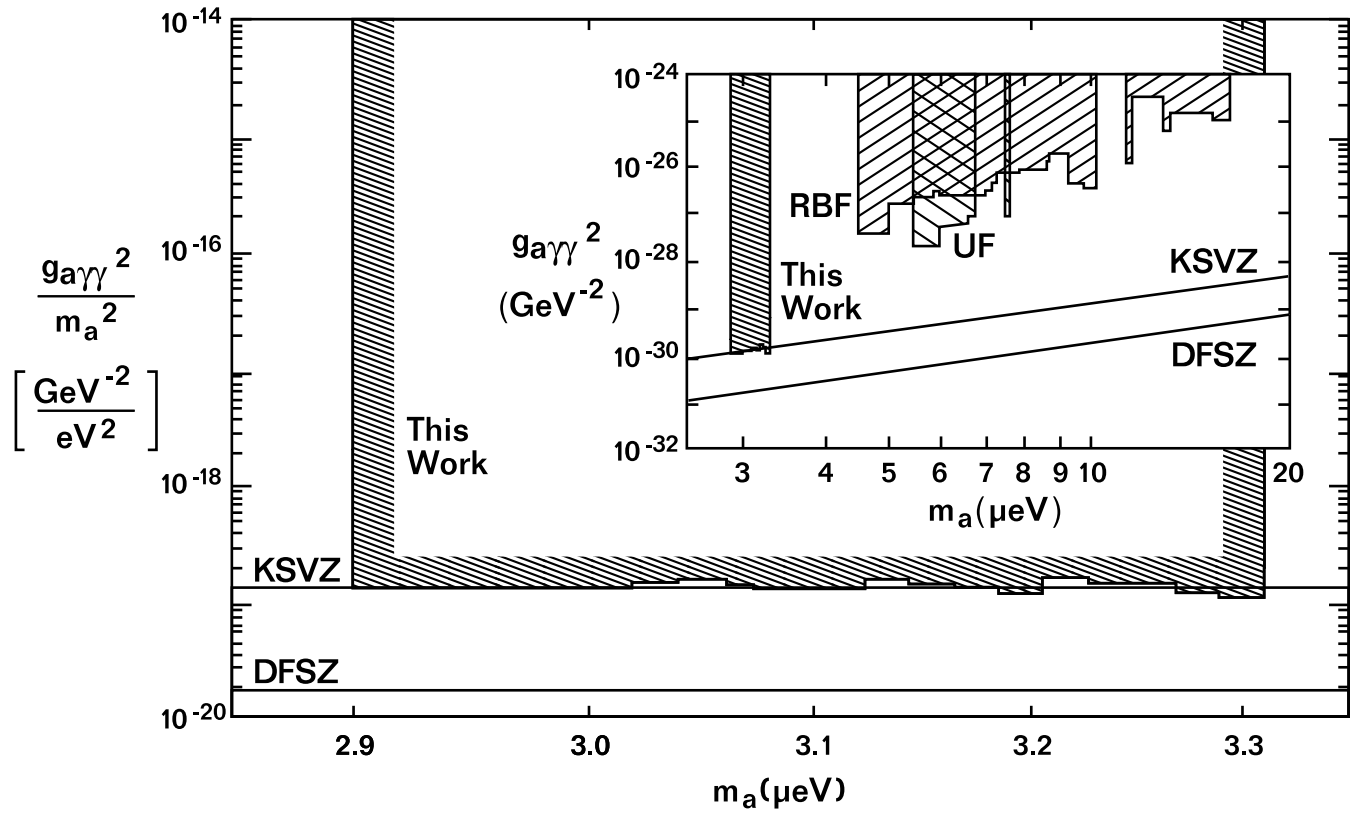
\begin{figure}
\caption{Axion couplings and
masses excluded at 90\% confidence by this experiment.
Also shown are KSVZ and DFSZ model predictions.
Indicated on the inset are the regions excluded by
earlier microwave cavity experiments; RBF
indicates Rochester-BNL-FNAL [11],
UF indicates University of Florida [12].
All results are scaled to $\rho_a=\rho_{halo}$.}
\label{limit}
\end{figure}

Figure\ \ref{limit}\ shows the axion couplings and
masses excluded at the 90\% confidence level by this analysis
for axions with the expected Maxwellian velocity distribution.
At 95\% (68\%) confidence level, the value of
$g_{a\gamma\gamma}^2/m_a^2$
excluded is approximately
1.6(1.0)$\times10^{-19}$GeV$^{-2}$/eV$^{2}$.
Also shown are KSVZ and DFSZ model predictions.
Indicated on the inset are the regions excluded by
earlier microwave cavity experiments\cite{experirbf,experiuf}.
The present experiment is more than two
orders-of-magnitude more sensitive, and is the
first to exclude a well-developed axion model (KSVZ)
at a realistic density over any mass range.
The significance of this result, however, is not
just the exclusion of a given axion model
over a narrow mass range at the most probable
local cold dark matter density (within the
full-width at half-maximum likelihood range of 
4.5--12.$\times10^{-25}$~g\,cm$^{-3}$\cite{Astro:Turner}).
Equally important, the sensitivity of the microwave
cavity scheme has been brought into the region of
interest, {\it i.e.,} where the axion
might plausibly be discovered.

The high-resolution channel is analyzed similarly.
The power spectra are binned on-line at resolutions
0.02~Hz, 0.16~Hz and 1.3~Hz, for which candidates
were defined as those peaks with more than
15$\sigma$, $8\sigma$ and $5\sigma$
excess power respectively.
After rescanning candidate peaks and eliminating
external noise peaks, no candidates remained.
This procedure resulted in upper power limits near
$3.3\times10^{-23}$~W (0.02~Hz/channel),
$5.0\times10^{-23}$~W (0.16~Hz/channel) and
$8.8\times10^{-23}$~W (1.3~Hz/channel).
We further searched for coincidences between medium- and
high-resolution candidates, and after excluding obvious
external RF sources, found none.

In conclusion, for the first time sufficient
sensitivity has been achieved to detect KSVZ
axions if they comprise the dark matter of
our own galactic halo.
Based upon our first results, we exclude at 90\% confidence
a KSVZ axion of mass between 2.9 and 3.3 $\mu$eV, assuming
halo axions have a Maxwellian velocity distribution.
If a significant fraction of halo axions are distributed
in a few narrow peaks, weaker axion two-photon
couplings are excluded.
Promising developments in amplifier and magnet technologies
may soon extend the sensitivity of the experiment by more
than an order of magnitude, permitting a search
for axions that couple more
weakly ({\it e.g.,} DFSZ) at lower halo densities.

The authors thank R. Bradley of NRAO for sharing his
amplifier expertise.
This research is supported by the U.S. Department
of Energy under contracts
W-7405-ENG-048,
DE-AC03-76SF00098,
DE-AC02-76CH03000,
DE-FC02-94ER40818,
DE-FG02-97ER41029 and
DE-FG02-90ER40560
and the National Science Foundation
grant number PHY-9501959.

\end{document}